\documentclass[conference,a4paper]{IEEEtran}

\ifCLASSINFOpdf
\else

\fi

\usepackage{graphicx}
\usepackage{epstopdf}
\usepackage{balance}
\usepackage{multicol}
\usepackage{url}
\usepackage{amsmath}
\usepackage{mathtools}
\usepackage{amssymb}
\usepackage{bbold}
\usepackage{esint}
\usepackage{amsfonts}
\usepackage{multirow}
\usepackage{wrapfig}
\usepackage{hyperref}
\usepackage{cite}
\usepackage{subcaption}
\usepackage{lipsum}
\usepackage{float}    
\usepackage{algorithm}
\usepackage[noend]{algpseudocode}
\usepackage{url}
\usepackage{caption}
\usepackage[none]{hyphenat}
\usepackage[usenames,dvipsnames]{color}

\begin{document}
\title{\LARGE DQN-Based Multi-User Power Allocation for Hybrid RF/VLC Networks}
\author{\IEEEauthorblockN{Bekir Sait Ciftler, Abdulmalik Alwarafy, Mohamed Abdallah, and Mounir Hamdi\\}
\IEEEauthorblockA{Division of Information and Computing Technology, College of Science and Engineering,\\
Hamad Bin Khalifa University, Doha, Qatar\\
\{bciftler, aalwarafy, moabdallah, mhamdi\}@hbku.edu.qa}}
\vspace{-5mm}
\maketitle 
\begin{abstract}
In this paper, a Deep Q-Network (DQN)
based multi-agent multi-user power allocation algorithm is proposed for hybrid networks composed of radio frequency (RF) and visible
light communication (VLC) access points (APs). The users are
capable of multihoming, which can bridge RF and VLC links for
accommodating their bandwidth requirements. By leveraging a
non-cooperative multi-agent DQN algorithm, where each AP is
an agent, an online power allocation strategy is developed to
optimize the transmit power for providing users' required data
rate. Our simulation results demonstrate that DQN's median
convergence time training is $90\%$ shorter than the Q-Learning
(QL) based algorithm. The DQN-based algorithm converges to
the desired user rate in half duration on average while converging
with the rate of $96.1\%$ compared to the QL-based algorithm's
convergence rate of $72.3\%.$ Additionally, thanks to its continuous
state-space definition, the DQN-based power allocation algorithm
provides average user data rates closer to the target rates than
the QL-based algorithm when it converges.
\end{abstract}
\vspace{-3mm}
\section{Introduction}
\label{sect:Introduction}

As the number of interconnected devices in our lives increases exponentially, spectrum scarcity becomes a bigger problem.
In recent years Visible Light Communication (VLC) has attracted attention due to its vast potential to provide high data rates and ubiquitous coverage indoors by utilizing visible spectrum.
VLC is based on the utilization of light-emitting diodes (LEDs), which are very energy-efficient and capable of exploiting unused visible spectrum\cite{haas2015hybrid,dimitrov2015principles,hammouda2018link}.
VLC has many advantages, such as cheap transmitters and receivers, low power consumption\cite{kashef2016energy}, and better physical layer security features~\cite{mostafa2014physical,al2019secrecy}.

However, VLC requires line-of-sight communication with a proper angle between the transmitter and the receiver.
Hence, its coverage area can be limited, even indoors.
As a consequence, VLC is usually used in hybrid systems along with the proprietary RF communication networks\cite{li2015cooperative}.
Hybrid RF/VLC systems are gaining popularity thanks to its many features, such as energy-efficiency, ubiquitous connectivity by utilizing existing infrastructure, and high throughput capacity.

The users of hybrid RF/VLC systems usually have the multihoming ability, which allows them to connect multiple APs simultaneously.
Power allocation in these networks is crucial in providing the necessary quality-of-service (QoS) for the applications while reducing the overconsumption of the power and possible interference with other network entities~\cite{li2018intelligent}.
The proper allocation of the transmit power is of great importance concerning varying channel conditions and user requirements.

Conventional power allocation mechanisms usually utilize optimization methods such as mathematical programming.
However, hybrid RF/VLC systems usually have complex models that result in intractable optimization problems for power allocation\cite{basnayaka2017design}.
Usually, the system model for power allocation requires approximations and relaxations on closed-form equations around the solution to provide satisfactory results.
Hence, the general method would be an approximation of the model for possible solutions \cite{ahmad2015survey}, or the hybrid RF/VLC network power allocation modeled with mixed-integer nonlinear programming as in \cite{li2015cooperative}.
Then, the model is simplified to a discrete linear programming problem by approximation around the solution. 
It is shown that the model becomes intractable as the number of parameters and network elements increases even with the simplified forms.

As the number of APs and user devices increases in these hybrid RF/VLC networks, these solutions become impractical with increased complexity. Therefore other techniques that do not rely on the complex system models are required.
Machine learning (ML) based solutions for power allocation have been gaining prominence due to their vast performance in solving the optimization problems without explicit system models and state transition dynamics\cite{hussain2020machine}.
Especially reinforcement learning (RL) based solutions are prevailing since they allow mapping of states and the best actions based on observations to maximize the defined cumulative reward.

\begin{figure}[t]
    \centering
    \includegraphics[width=.7\linewidth]{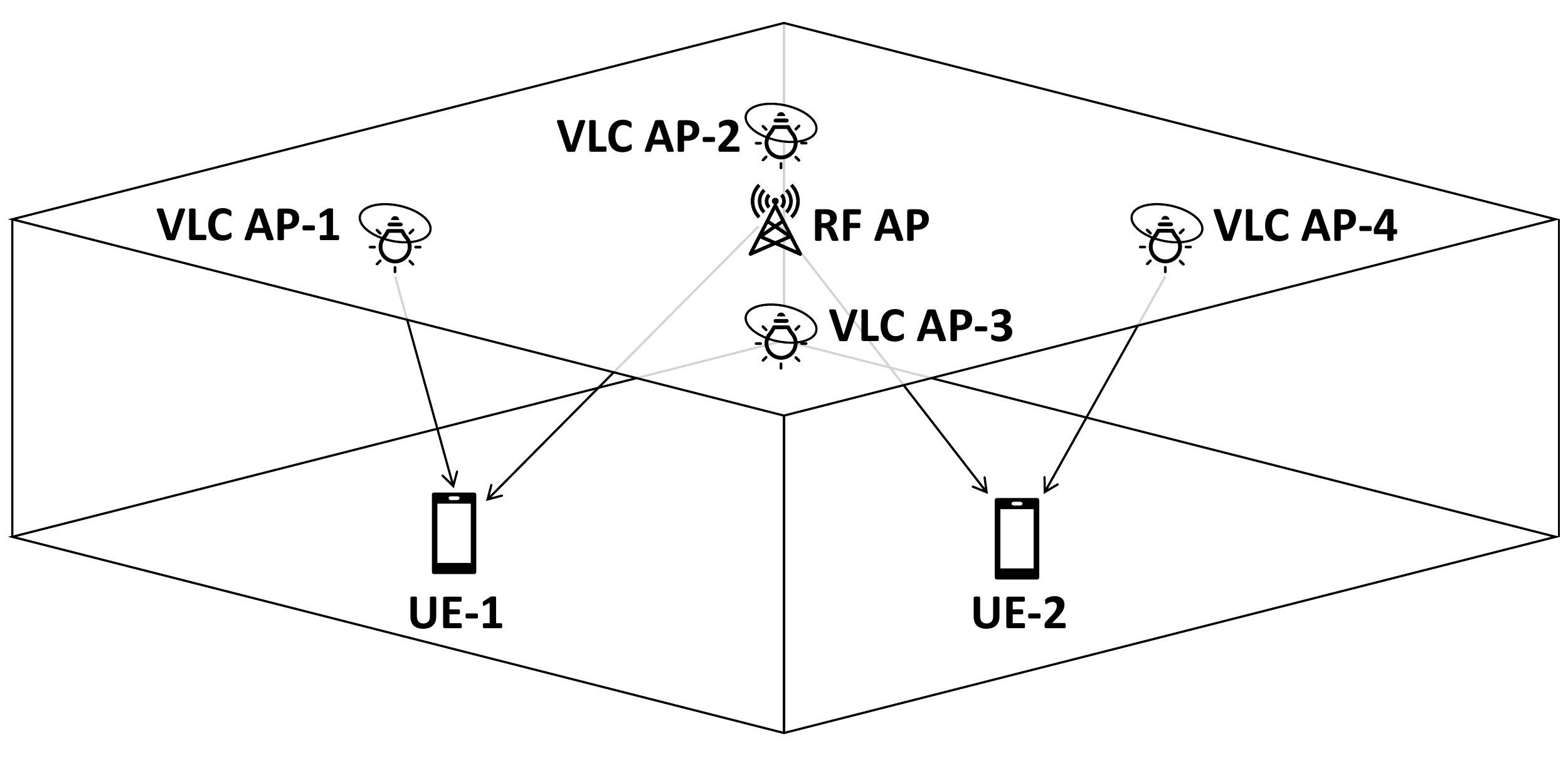}
    \caption{Hybrid RF/VLC Communication System.}
    \label{fig:SystemModel}
    \vspace{-5mm}
\end{figure}

In our work, we propose a multi-agent DQN-based multi-user power allocation scheme for hybrid RF/VLC networks, as shown in Fig.~\ref{fig:SystemModel}.
The power allocation problem is defined as an optimization problem in Section~\ref{sect:SystemModel} to provide the necessary data rate to users while minimizing power consumption.
The proposed methodology allows having a continuous state-space where the gap between the target rates and the actual rates of users can be observed precisely.
Hence the actions for allocation of power will be more efficient and precise as well.
Our simulation results show that the proposed method converges $10$ faster than the Q-Learning (QL) based algorithm.
Additionally, the proposed method achieves a closer average user rate to the target rate thanks to its continuous state-space.

Our main contributions in this work are listed below:
\begin{itemize}
    \item We propose and implement a non-cooperative multi-agent DQN-based algorithm to solve the multi-user power allocation problem of hybrid RF/VLC systems with the continuous state-space definition.
    \item We define a precise reward function for the stability of non-cooperative RL-based algorithms for multi-user power allocation.
    \item We benchmark convergence time for multi-agent QL-based and DQN-based power allocation algorithms.
    \item We show that DQN outperforms a QL-based algorithm with shorter convergence times.
\end{itemize}

This paper is structured as follows.
In Section~\ref{sect:RelatedWork}, a brief overview of existing literature on power allocation for hybrid RF/VLC networks is presented.
We provide the system model for the hybrid RF/VLC communication system in Section~\ref{sect:SystemModel}.
Subsequently, we explain the QL-based and DQN-based multi-user power allocation algorithms in Section~\ref{sect:RLBasedmulti-user}.
Numerical results for the provided techniques are given in Section~\ref{sect:Results}.
Finally, in Section~\ref{sect:Conclusion}, we present our future work and concluding remarks.

\section{Literature Review}
\label{sect:RelatedWork}

Hybrid RF/VLC systems are a hot prospect for energy-efficient and ubiquitous wireless communications.
This section presents a brief literature review on the power allocation for hybrid RF/VLC networks.

All of the conventional optimization techniques for the VLC network performance is presented as a survey in \cite{obeed2019optimizing}.
The VLC systems and channel models are provided in-depth for optimization algorithms.
Resource and power control with AP assignment is reviewed in detail.
The optimization techniques proposed are conventional techniques that are model-based and require full observation of the channel.


As an example of the conventional methods for resource allocation for hybrid RF/VLC systems, the authors investigated the cell formation and frequency reuse patterns in the context of load balancing in \cite{li2015cooperative}.
The performance of this hybrid VLC system is proven to be providing high area spectral efficiency, and using a hybrid configuration allows to provide the highest grade of fairness in most of the scenarios.
In \cite{li2015cooperative}, the hybrid system resource allocation modeled with mixed-integer nonlinear programming (MINLP); however, it is simplified to discrete linear programming by approximation. It is shown that the model becomes intractable as the number of elements increases.

A QL-based power allocation technique is proposed for hybrid RF/VLC networks in a distributed fashion in \cite{Kong2019TwoTimescale}.
In this study, a multi-agent QL-based technique is proposed where each AP is an independent agent that interacts with the environment in a two-timescale power allocation scheme.
The proposed methodology satisfies the QoS requirements for the users on average.
However, classical solutions such as QL have problems with scalability and limited mapping of observations due to the state space's discrete definition.

The energy-efficient resource of software-defined hybrid RF/VLC systems is studied in \cite{zhang2018energy}.
The authors developed an optimization framework that considers backhaul constraints, QoS requirements, energy-efficiency, and inter-cell interference limits for a heterogeneous VLC and RF small cell network.
The formulated optimization problem is solved by the alternative direction method of multipliers (ADMM) method.
The simulation results show that the proposed scheme can converge within a few iterations while increasing the throughput significantly and avoiding interference by limiting power consumption.


All of the references mentioned above utilize either conventional optimization techniques or RL techniques with discrete state-space. In our work, we propose a multi-agent DQN-based multi-user power allocation algorithm for hybrid RF/VLC systems to allocate power more precisely, considering users' data rate requirements and their actual rates.

\section{System Model and Problem Formulation}
\label{sect:SystemModel}
We consider a multi-user downlink resource allocation problem for multihoming hybrid RF/VLC networks in this work.
Our system model consists of a single RF AP and $K$ VLC APs.
There are $N$ mobile users with RF and VLC receivers with multihoming capability.
At timestep $t$, the channel gain for the link between user $u$ and VLC AP $l$ can be represented as\cite{basnayaka2017design}:
\begin{align} G_{\mathrm {VLC}}^{(u,l)}(t)=&\frac {(m+1)A_{pd}\lambda \cos ^{m}(\theta _{tx}^{(u,l)}(t))}{2\pi \left ({\left ({x^{(u,l)}(t) }\right)^{2} + y^{2} }\right) } \\&\times H_{f}(\theta _{rx}^{(u,l)}(t))H_{c}(\theta _{rx}^{(u,l)}(t)) \cos (\theta _{rx}^{(u,l)}(t)),\nonumber\end{align}
where $x$ and $y$ are the horizontal and the vertical distances between the user and the VLC AP.
$A_{pd}$ is the photodiode (PD) effective detection area, $\lambda$ is PD responsivity, and $\theta_{tx}$ and $\theta_{rx}$ represents angle of irradiance and angle of incidence, respectively\cite{basnayaka2017design}.
The PDs of all users are assumed to be facing vertically upwards for simplicity (i.e. $\theta^{(u,l)}_{tx}=\theta^{(u,l)}_{rx}$).
The gain of the user's optical filter and the gain of the optical concentrator is represented by $H_f(\theta^{(u,l)}_{rx}(t))$ and $H_c(\theta^{(u,l)}_{rx}(t))$, respectively.
Additionally, $m=-1/\log_2(\cos(\Psi_{1/2}))$, where $\Psi_{1/2}$ is the semi-angle at half-power of the LED.
The gain of the user's optical filter is assumed to be $1$ throughout this manuscript\cite{Kong2019TwoTimescale}, while the optical concentrator's gain is given in below equation:
\begin{equation} H_{c}(\theta _{rx}^{(u,l)}(t)) = \frac {n_{c}^{2}}{\sin ^{2}(\Psi _{\mathrm {fov}})} {\mathbb{1}} \left ({0\leq \theta _{rx}^{(u,l)}(t)\leq \Psi _{\mathrm {fov}} }\right),\end{equation}
where $\Psi_{fov}$ stands for the half of the PD's field-of-view (fov), $\mathbb{1}(\cdot)$ is the indicator function, and $n_c$ is the optical concentrator's reflective index\cite{basnayaka2017design}.


The coverage regions of the VLC APs are exclusive since each VLC AP allocates orthogonal frequencies for bandwidth.
Each AP's total bandwidth is divided equally for the total number of users within its coverage area.
The transmit power for each VLC AP is determined by a centralized entity at the beginning of each timestep $t$.
The achieavable rate of the link between the user $u$ and the VLC AP $l$ in timestep $t$ represented with:
\begin{equation} \small \!\!\!R_{\mathrm {VLC}}^{(u,l)}(t) = \frac {W_{\mathrm {VLC}}}{2} \log _{2}\left ({1 {+} \frac {\left ({\kappa m_{d} P_{\mathrm {VLC}}^{(u,l)}(t_{n}) G_{\mathrm {VLC}}^{(u,l)}(t)}\right)^{2} }{ W_{\mathrm {VLC}} \sigma ^{2}_{\mathrm {VLC}}}}\right),\label{eq:4}\end{equation}
where $W_{VLC}$ is the VLC link bandwidth, $\kappa$ is the optical to electric conversion efficiency, $m_d$ is the modulation depth. $\sigma^2_{VLC}$ represents the noise power spectral density (PSD) of VLC links.
$P^{(u,l)}_{VLC}(t)$ is the optical (transmit) power of VLC AP $l$ for user $u$ in timestep $t$\cite{basnayaka2017design}.

The gain of the RF link in timestep $t$ is defined as below:
\begin{equation} G_{\mathrm {RF}}^{(u)}(t) = 10^{- L(d_{\mathrm {RF}}^{(u)}(t))/10 } | h_{\mathrm {RF}}^{(u)}(t) |^{2},\end{equation}
where $h^u_{RF}(t)$ stands for the small-scale fading, modeled with an exponential random variable with mean $2.46$ dB. 
$L(d)$ is the path loss component defined as in below equation:
\begin{equation}
    L(d)=47.9+10\nu\log_{10}(d/d_0)+X \textrm{ (dB)},
\end{equation}
where $d$ indicates the distance between the transmitter and the receiver, $d_0=1$~m, $\nu=1.6$, $X$ is a Gaussian random variable with its mean equal to zero with standard variance of $1.8$ dB representing the shadowing component\cite{Kong2019TwoTimescale}.

The achievable rate at the user $u$ from the RF link in timestep $t$ is given as:
\begin{equation} R_{\mathrm {RF}}^{(u)}(t) = W_{\mathrm {RF}} \log _{2}\left ({1 + \frac {P_{\mathrm {RF}}^{(u)}(t) G_{\mathrm {RF}}^{(u)}(t) }{ W_{\mathrm {RF}} \sigma ^{2}_{\mathrm {RF}}}}\right),\end{equation}
where $W_{RF}$ is the bandwidth of each RF link, $P^{u}_{RF}(t)$ is the allocated transmit power for the link between user $u$ and RF AP in timestep $t$.
The power spectral density of additive white Gaussian noise (AWGN) for RF links is represented with $\sigma^2_{RF}$.
Since users are assumed to have multihoming capability, the RF and VLC links' simultaneous use are possible.
Hence the total achieveable rate for the user $u$ at timestep $t$ becomes the total actual rate of both links:
\begin{equation}
R^{(u)}(t)=R^{(u)}_{RF}(t)+R^{(u,l)}_{VLC}(t),   
\end{equation}
where $l$ is the associated VLC AP for user $u$ at timestep $t$.
Our goal in this paper is to develop a DRL-based solution to control the transmit powers to achieve maximum utility function value based on each user's actual rate.

The optimization problem for adjusting the transmit powers accordingly defined as follows
\begin{align} \underset {\{ P_{\mathrm {RF}}^{(u)}(t) \}, \{ P_{\mathrm {VLC}}^{(u,l)}(t) \}}{\max}&\sum _{t=1}^{\infty }~U(t) \label{eq:optmax} \\ \mathrm {s.t.} \qquad ~~&\sum _{u=1}^{U} P_{\mathrm {RF}}^{(u)}(t) \leq P_{\mathrm {RF}}^{\mathrm {max}} 
\nonumber \\&\sum _{u=1}^{U} P_{\mathrm {VLC}}^{(u,l)}(t) \leq P_{\mathrm {VLC}}^{\mathrm {max}}~~ \forall l,\nonumber\\
&P^{(u,l)}_{VLC}\geq0,~P^{(u)}_{RF}\geq0 ~~\forall u,l\nonumber
\end{align}
where $U(t)$ is the utility function for the timestep $t$.
In our proposed scheme, the utility function defined as below:
\begin{align}
    U(t)=\sum_{u=1}^{N} B^{(u)} -|R^{(u)}(t)-T^{(u)}|,\label{eq:utility}
\end{align}
where $B^{(u)}$ is the target rate band which defines the vicinity of $T^{(u)}$, which is the target rate requirement for QoS to be provided to the user $u$.
In our simulations, we defined the target band as follows
\begin{align}
    B^{(u)}=\max\{0.05 \times T^{(u)}, 0.5\}.
\end{align}
This band's definition provides the system's stability since there might not be a possible solution for a particular target rate due to discrete power levels.
For the sake of simplicity, we assumed the target rates of users are static within an episode.

\section{RL-based Multi-User Multi-Agent Power Allocation}
\label{sect:RLBasedmulti-user}
In this section, we present two different, QL-based and DQN-based power allocation methods in which the transmit powers of the RF AP and the VLC APs are adjusted in every timestep to optimize the downlink data rates of users.
In both algorithms, separate agents are not in communication with each other; hence they work non-cooperatively.

The state space for the system is defined by the target rate and the actual rate of the users as follows
\begin{align}
    \textbf{s}_t = [s^{(1)}_t,~\cdots,~s^{(u)}_t,~\cdots,~ s^{(N)}_t],\label{eq:statespace}
\end{align}
where  separate state-space entries for each user is defined separately since QL is bounded by a state-action table, however we have continuous state space for DQN.

\begin{algorithm}[t]
\caption{QL-based Power Allocation}
    \label{Alg:QL}
    \small{
\begin{algorithmic}[1]
\State \textbf{Initialization:} Set $t=0$. Initialize Q-values for all state-action pairs as $Q_{VLC}^{(l)}(\textbf{s},\textbf{a})=0$ for VLC APs, and $Q_{RF}(\textbf{s},\textbf{a})=0$ for RF AP.
\For{$t=1$ to $\infty$}
\State Observe state $\textbf{s}_t$. 
    \For {$l=1$ to $K$}
    \State Generate a random number $x$ from $[0,1]$.
    \If{$x\leq\epsilon(t)$} Select a random action $\boldsymbol{a}^{(l)}_t$ from action space of VLC AP $(A_{VLC})$.
    \Else{} Select $\textbf{a}^{(l)}_t$ that gives the largest Q-value according to $\arg\max\limits_{a^{(l)} \in A_{VLC}} Q_{VLC}^{(l)}(\textbf{s}_t,\textbf{a}^{(l)})$
    \EndIf
    \State \textbf{end if}
    \EndFor
    \State \textbf{end for}
    \State Generate a random number $x$ from $[0,1]$.
    \If{$x\leq\epsilon(t)$} Select a random action $\textbf{a}_{RF}$ from action space of RF AP $(A_{RF})$.
    \Else{} Select $\textbf{a}^{RF}_t$ that gives the largest Q-value according to $\arg\max\limits_{a^{RF} \in A_{RF}} Q_{RF}(\textbf{s}_t,\textbf{a}^{RF})$
    \EndIf
    \State \textbf{end if}
    \State Execute all actions $\textbf{a}^{(l)}_t$ at VLC APs $l=1$ to $K$, and $\textbf{a}^{RF}_t$ at RF AP.
    \State Receive the rewards $r_t$ using \eqref{eq:reward}
    \State Observe the new state $\textbf{a}_{t+1}$ using \eqref{eq:stateQ}
    \State Update $Q_{VLC}^{(l)}$ for VLC APs and $Q_{RF}$ for RF as follows
    \Statex $Q_{VLC}^{(l)} (\textbf{s}_t,\textbf{a}^{(l)})\leftarrow (1-\alpha)~Q_{VLC}^{(l)} (\textbf{s}_t,\textbf{a}^{(l)}) + \alpha \bigg( r_t + \gamma \max\limits_{\textbf{a}^{(l)} \in A_{VLC}} Q_{VLC}^{(l)}(\textbf{s}_{t+1},\textbf{a}^{(l)})\bigg)$
    \Statex $Q_{RF} (\textbf{s}_t,\textbf{a}^{RF})\leftarrow (1-\alpha)~Q_{RF} (\textbf{s}_t,\textbf{a}^{RF}) + \alpha \bigg( r_t + \gamma \max\limits_{\textbf{a}^{RF} \in A_{RF}} Q_{RF}(\textbf{s}_{t+1},\textbf{a}^{RF})\bigg)$
\EndFor
\State \textbf{end for}
\end{algorithmic}}
\end{algorithm}

\subsection{QL-based Power Allocation}

In this subsection, a QL-based power allocation method in which the power allocations of VLC APs and RF AP are individual agents utilizing QL to learn optimal power allocation to provide necessary target rates $T^{(u)}$ to users by taking the current status of users as explained in Algorithm~\ref{Alg:QL}.

\subsubsection{State-space} The state space for the QL-based power allocation is based on each user's actual rate and target rate as follows
\begin{equation}
  s^{(u)}_t =
  \begin{cases}
    1, & \text{if $R^{(u)}(t) < T^{(u)}$} \\
    2, & \text{if $R^{(u)}(t) > T^{(u)} + B^{(u)} $} \\
    3, & \text{if $T^{(u)} + B^{(u)} \geq R^{(u)}(t) \geq T^{(u)}$}
  \end{cases},\label{eq:stateQ}
\end{equation}
where $1$ means the actual rate of the user is below target rate, $2$ means the actual rate is above the target rate much more than the target band, while $3$ means the user's actual rate is within targeted band.
The state-space is the same for all the agents, independent of users' location and agents' actions.

\subsubsection{Action-space}
\label{subsect:QLActionSpace}
The action-space of each agent in the system is defined as the power level of the APs either VLC or RF.
Hence the sets of transmit powers at VLC AP or RF AP can be defined as
\begin{align}
    \mathcal{P}_{VLC}=\{P_{VLC,1},P_{VLC,2},\dots,P_{VLC,V_P}\},\label{eq:pvlc}
\end{align}
and
\begin{align}
    \mathcal{P}_{RF}=\{P_{RF,1},\dots, P_{RF,k},\dots,P_{RF,R_P}\},
\end{align}
where $V_P$ and $R_P$ refers to the number of power levels at VLC APs and RF AP, respectively.
The action space of a VLC AP can be expressed as
\begin{align}
    A_{VLC} = \{\textbf{a}_1, \dots, \textbf{a}_i, \dots, \textbf{a}_{V_A}\}
\end{align}
where $\textbf{a}_i = [a_i^{(1)}, \cdots, a_i^{(u)}, \cdots, a_i^{(U)}]$ is a vector with size $U$ (i.e. number of users), where $a_i^{(u)}\in \mathcal{P}_{VLC}$ refers to transmit power levels allocated to users, which is bounded by the below equation due to power constraint
\begin{align}
    \sum_{u=1}^{U}a_i^{(u)} \leq P^{max}_{VLC}.
\end{align}
As a consequence, there are possible $V_A$ possible transmit power combinations according to above equation.
Similarly, RF AP agent has the below action space with the size of $R_A$ possible combinations 
\begin{align}
    A_{RF} = \{\textbf{a}_1, \dots, \textbf{a}_i, \dots, \textbf{a}_{R_A}\}
\end{align}
where $\textbf{a}_i = [a_i^{(1)}, \cdots, a_i^{(u)}, \cdots, a_i^{(U)}]$ is a vector with size $U$ (i.e. number of users), where $a_i^{(u)}\in \mathcal{P}_{RF}$ refers to transmit power levels allocated to users, which is bounded by the below equation due to power constraint
\begin{align}
    \sum_{u=1}^{U}a_i^{(u)} \leq P^{max}_{RF}.\label{eq:pmaxrf}
\end{align}
\subsubsection{Reward function}
\label{subsect:QLReward}
The optimization problem defined in \eqref{eq:optmax} and \eqref{eq:utility} aims to minimize the difference between the actual rate and the target rate of users.
In our work, we define the reward function using \eqref{eq:utility}, hence the reward is as follows
\begin{align}
    r_t = \sum_{u=1}^{N} B^{(u)} -|R^{(u)}(t)-T^{(u)}|. \label{eq:reward}
\end{align}

\subsubsection{Exploration vs. Exploitation}

The exploration vs. exploitation trade-off is one of the critical success factors in RL-based systems.
Our algorithm has used a time-dependent $\epsilon$-greedy technique to balance exploration and exploitation.
The epsilon function ($\epsilon(t)$) for the algorithm is defined as follows
\begin{equation}
  \epsilon(t) =
  \begin{cases}
    0.99^{(t-1)}, & \text{if $0.99^{(t-1)}>0.1$} \\
    0.1, & \text{if $0.99^{(t-1)}\leq0.1$}
  \end{cases}.\label{eq:epsilonT}
\end{equation}

\begin{algorithm}[t]
\caption{DQN-based Power Allocation}
\label{Alg:DQN}
\small{
\begin{algorithmic}[1]
\State \textbf{Initialization:} Initialize replay memory of VLC AP agents $\mathcal{D}^{(l)}_{VLC}$ and RF AP agent $\mathcal{D}_{RF}$ with capacity $M$.
\State Initialize action-value functions $Q^{(l)}_{VLC}$ and $Q_{RF}$ with random weights $\theta^{(l)}_{VLC}$ and $\theta_{RF}$.
\For {$t=1$ to $\infty$}
\State Initialize the state $\textbf{s}_t$ with initial observation
\For {$l=1$ to $K$}
\State Generate a random number $x$ from $[0,1]$
\If{$x\leq\epsilon(t)$} Select a random action $\textbf{a}^{(l)}_t$ from action space of VLC AP $A_{VLC}$.
\Else{} Select $\textbf{a}^{(l)}_t$ that gives the largest Q-value according to $\arg\max\limits_{\textbf{a}^{(l)} \in A_{VLC}} Q^{(l)}_{VLC}(\textbf{s}_t,\textbf{a}^{(l)};\theta^{(l)}_{VLC})$
\EndIf
\State \textbf{end if}
\EndFor
\State \textbf{end for}
\State Generate a random number $x$ from $[0,1]$.
\If{$x\leq\epsilon(t)$} Select a random action $\textbf{a}^{RF}_t$ from action space of RF AP $A_{RF}$.
\Else{} Select $\textbf{a}^{RF}_t$ that gives the largest Q-value according to $\arg\max\limits_{\textbf{a}^{RF} \in A_{RF}} Q_{RF}(\textbf{s}_t,\textbf{a}^{RF};\theta_{RF})$
\EndIf
\State \textbf{end if}
\State Execute all actions $\textbf{a}^{(l)}_t$ at VLC APs $l=1$ to $K$, and $\textbf{a}^{RF}_t$ at RF AP.
\State Receive the reward $r_t$ according to \eqref{eq:reward} and observe the new state $\textbf{s}_{t+1}$ according to \eqref{eq:StateDQN} 
\State Store transition $(\textbf{s}_t,\textbf{a}^{(l)}_t,r_t,\textbf{s}_{t+1})$ for VLC APs in $\mathcal{D}^{(l)}_{VLC}$ and $(\textbf{s}_t,\textbf{a}^{RF}_t,r_t,\textbf{s}_{t+1})$ for RF AP in $\mathcal{D}_{RF}$
\For{$l=1$ to $K$}
\State Sample random minibatch transitions from $D^{(l)}_{VLC}$
\State Set $y_j = r_j + \gamma \max\limits_{\textbf{a}^{(l)}}Q^{(l)}_{VLC}(\textbf{s}_{j+1},\textbf{a}^{(l)};\theta^{(l)}_{VLC})$
\State Set $\hat{y}_j = Q^{(l)}_{VLC}(\textbf{s}_j,\textbf{a}^{(l)}_j;\theta^{(l)}_{VLC})$
\State Update the weights with gradient descent:
\Statex $~  ~  ~ ~ ~ ~ ~ ~ ~  \theta^{(l)}_{VLC}\leftarrow\theta^{(l)}_{VLC}+\alpha\nabla\frac{1}{2}(y_j-\hat{y}_j)^2$
\EndFor
\State \textbf{end for}
\State Sample random minibatch transitions from $\mathcal{D}_{RF}$
\State Set $y_j = r_j + \gamma \max\limits_{\textbf{a}^{RF}}Q_{RF}(\textbf{s}_{j+1},\textbf{a}^{RF};\theta_{RF})$
\State Set $\hat{y}_j = Q_{RF}(\textbf{s}_j,\textbf{a}^{RF}_j;\theta_{RF})$
\State Update the weights with gradient descent:
\Statex $~ ~ ~ ~ ~ \theta_{RF}\leftarrow\theta_{RF}+\alpha\nabla\frac{1}{2}(y_j-\hat{y}_j)^2$
\EndFor
\State \textbf{end for}
\end{algorithmic}}
\end{algorithm}

\subsection{DQN-based Power Allocation}
In this subsection, we propose a DQN-based power allocation method that utilizes continuous state-space of DQN to alleviate the information on the actual rate and users' target rate for shortening convergence time and using power more efficiently.
In this method,  RF AP and each VLC AP acts as a separate agent without coordination.
Hence, they can only observe the users' actual and target rates and whether they are within their coverage area.
The algorithm for multi-agent DQN-based power allocation is provided in Algorithm~\ref{Alg:DQN}.

\subsubsection{State-space}
The state space of the DQN agents are defined using \eqref{eq:stateQ} as follows
\begin{align}
    s^{(u)}_t = [R^{(u)}(t),~T^{(u)}]^T,\label{eq:StateDQN}
\end{align}
where $R^{(u)}(t)$ is the actual and $T^{(u)}$ rate of the user $u$.
This state-space definition allows our agents to act on the actual difference and learn to be more clinical to get closer to the target rate.

\subsubsection{Action-space}
The action space definition is same as QL action space as given in \eqref{eq:pvlc}-\eqref{eq:pmaxrf}.
\subsubsection{Reward function}
The reward function definition is the same as QL agents, as given in \eqref{eq:reward}.
\subsubsection{Exploration vs. Exploitation}
The epsilon function ($\epsilon(t)$) for the algorithm is defined in \eqref{eq:epsilonT}.

\section{Numerical Results}
\label{sect:Results}

\begin{table}[t]
\centering
\caption{Simulation parameters for the hybrid network.}
\label{table:simparameters}
\begin{tabular}{|l|c|}
\hline
\multicolumn{1}{|c|}{\textbf{Parameter}} & \multicolumn{1}{c|}{\textbf{Value}} \\ \hline
Maximum transmit power for RF links ($P^{max}_{RF}$) & $0.01$ W \\ \hline
PSD of AWGN at the RF Links ($\sigma^2_{RF}$) & $-57$ dBm/MHz \\ \hline
Bandwidth for RF Links ($W_{RF}$) & 5 MHz \\ \hline
Maximum transmit power for VLC links ($P_{VLC}^{max}$) & 2 W \\ \hline
PSD of noise in the VLC links ($\sigma^2_{VLC}$) & $-100$ dBm/MHz \\ \hline
Total bandwidth for each VLC AP ($W_{VLC}$) & $20$ MHz\\ \hline
The height of the ceiling ($y$) & $3$ meters\\ \hline
Half of the PD's field-of-view ($\Psi_{fov}$) & $45^{\circ}$ \\ \hline
The semi-angle at half power of the LED ($\Psi_{1/2}$) & $60^{\circ}$ \\ \hline
The effective detection area of the PD ($A_{pd}$) & $10^{-4}$\\ \hline
Responsivity of the PD ($\lambda$) & $0.4$ \\ \hline
The gain of the optical filter ($H_f$) & $1$\\ \hline
The reflective index of the optical concentrator ($n_c$) & $1.5$\\ \hline
Optical to electric conversion efficiency ($\kappa$) & $1$\\ \hline
Number of VLC APs ($K$) & $4$\\ \hline
Number of RF AP ($K_{RF}$) & $1$\\ \hline
Number of UEs ($N$) & $2$\\ \hline
Learning Rate ($\alpha$) & 0.5\\\hline
Discount Factor ($\gamma$) & 0.5\\\hline
\end{tabular}
\end{table}

We consider a $12$ m $\times$ $12$ m room, with the ceiling height of $3 m$. The room's center is the origin point $(0,0)$, and RF AP is located at the origin.
Four VLC APs are located at $(-3,-3),(-3,3),(3,-3)$ and $(3,3)$.
We have simulated the given system in $1000$ Monte Carlo experiments where two users are randomly placed with $x$ and $y$ coordinates uniformly distributed within the room dimensions.
We have executed our simulations until the convergence is achieved, defined as having average user rates within target bands for all UEs for at least $100$ iterations.
The rest of the system parameters for the simulations are provided in Table~\ref{table:simparameters}.
TensorFlow and Keras libraries are used to implement the neural networks for DQN agents.
Each agent's DQN has $3$ hidden layers with $32$ nodes in each of them.
The loss function used is mean square error, and the optimizer is \emph{Adam} optimizer.

\begin{figure}[t]
    \centering
    \includegraphics[width=0.81\linewidth]{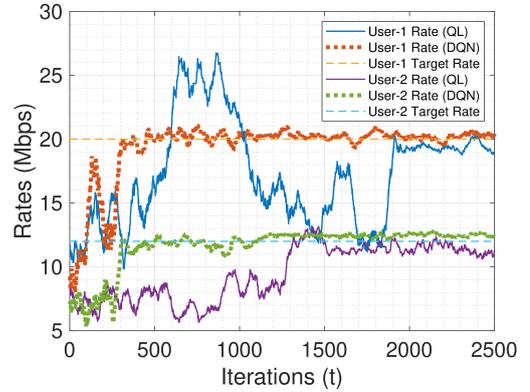}
    \caption{User Rate (Mpbs) vs Iterations (t) where user target rates are 20 Mbps and 12 Mbps.}
    \label{fig:TargetRates1}
    \vspace{-5mm}
\end{figure}

As a sample implementation, in Fig.~\ref{fig:TargetRates1}, we evaluate the two algorithms considering the actual user rates as a benchmark for the target rates of $20$ Mbps and $12$ Mbps for User-1 and User-2, respectively.
Both algorithms initially begin with lower transmit powers and take different actions within time.
In Fig.~\ref{fig:TargetRates1}, we can observe that the DQN-based power allocation algorithm achieves convergence within $306$ iterations by achieving the target rate for both users, whereas the QL-based power allocation algorithm requires nearly $2000$ iterations to reach a feasible solution with the desired average rate for both users.
The QL-based power allocation algorithm reaches the target rate for User-1, around $600$ iterations, and User-2 $1300$ iterations separately. However, it is not the desired state due to the other user's rate in each situation, which is much lower than its target rate.
Since the power is shared between users, it requires around $1900$ iterations for the QL-based algorithm to reach a feasible solution and convergence.
Another observation on Fig.~\ref{fig:TargetRates1}, can be made regarding the difference between the actual and the users' target rates.
Thanks to its continuous state space definition, the DQN-based algorithm converges to the target band in a shorter number of iterations and a closer value to the target rate than the QL-based algorithm.

\begin{figure}[t]
    \centering
    \includegraphics[width=0.81\linewidth]{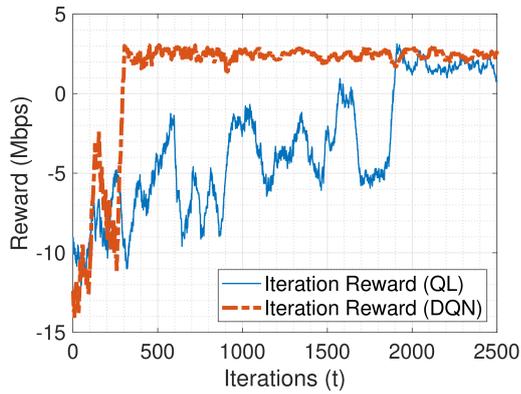}
    \caption{Reward comparison of QL and DQN Algorithms.}
    \label{fig:Rewards}
    \vspace{-5mm}
\end{figure}

In Fig.~\ref{fig:Rewards}, reward for each iteration for QL and DQN agents are provided considering the sample case of Fig.\ref{fig:TargetRates1}.
Note that the reward for each agent in the system is the same due to the reward function definition in \eqref{eq:reward}.
This result shows that convergence speed, performance, and the DQN-based algorithm are much better than the QL-based algorithm.
DQN-based algorithm reaches convergence in $300$ iterations, where the QL-based algorithm requires more than $1900$ iterations, and the DQN-based algorithm converges to a better result, with larger reward.

\begin{figure}[t]
    \centering
    \includegraphics[width=0.81\linewidth]{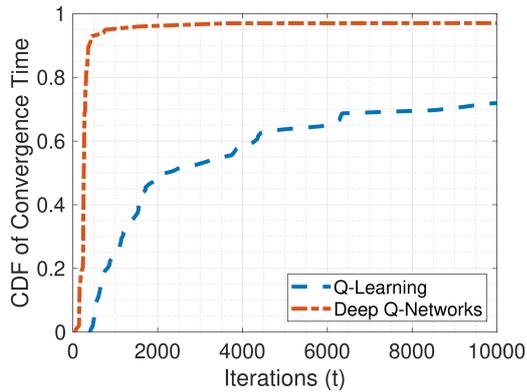}
    \caption{Convergence CDFs of QL and DQN.}
    \label{fig:ConvergenceTime}
    \vspace{-5mm}
\end{figure}

The convergence time CDFs of the two algorithms are provided in Fig.~\ref{fig:ConvergenceTime} over $1000$ Monte-Carlo experiments where users are distributed over the simulation area uniformly in each experiment.
QL-based power allocation algorithm has a median convergence time of $1989$ iterations, while the DQN-based power allocation algorithm has a median convergence time of $203$ iterations.
This result shows us using a continuous state space allows the DQN-based algorithm to converge $10$ times faster than the QL-based algorithm.
Additionally, we observe that the DQN-based power allocation scheme converges to the desired state (stability within target band) above $96.1\%$ of the time, while the QL-based algorithm could reach convergence $72.3\%$ of the time.

\section{Conclusion}
\label{sect:Conclusion}

This paper investigates the power allocation for hybrid RF/VLC networks with multiple users.
A multi-agent DQN-based algorithm is proposed to take precise actions considering the difference in users' actual rates and target rates.
It is shown that the DQN-based algorithm's median convergence time is $90\%$ shorter compared to the QL-based algorithm. Additionally, the DQN-based algorithm's performance on closing the gap between users' actual rates and target rates is better than the QL-based algorithm.
In our future work, our goal is to develop a Deep Deterministic Policy Gradient (DDPG) based power allocation algorithm to have a continuous action space to handle power allocation more precisely for converging to the exact target rates instead of target bands.
\bibliographystyle{IEEEtran}
\bibliography{refs}
\end{document}